# Representation of dynamic spatial phenomena:

# The daily car accessibility in London

MOYA-GÓMEZ, BORJA (*)

Transport, Infrastructure and Territory Research Group (t-GIS)

Human Geography Department

Faculty of Geography and History

Universidad Complutense de Madrid (UCM)

c/ Profesor Aranguren S/N, 28040 Madrid, Spain

Phone: +34 91 394 57 51

e-mail: bmoyagomez@ucm.es

GARCÍA-PALOMARES, JUAN CARLOS

Transport, Infrastructure and Territory Research Group (t-GIS)

Human Geography Department

Faculty of Geography and History

Universidad Complutense de Madrid (UCM)

c/ Profesor Aranguren S/N, 28040 Madrid, Spain

Phone: +34 91 394 59 52

e-mail: jcgarcia@ghis.ucm.es

* Corresponding author

**Abstract**

The map presented in this paper shows the effect of congestion on daily accessibility in the London metropolitan area. Because of its dynamic nature, it is challenging to both calculate the effects of this phenomenon and to represent it clearly on simple maps. Although we can use many traditional techniques for this purpose, they are usually static, and they cannot easily provide a comprehensive map of all the effects studied. In this paper, we have represented our results by two cartographic techniques rarely used in accessibility studies - cartograms and 3D maps, which we believe can achieve a more striking representation in animations of both the traffic-induced spatial distortion and the accessibility levels obtained. These animated maps show the direct space-time link between congestion and accessibility, and can therefore give a more detailed overview of the consequences of this phenomenon.

## 1. INTRODUCTION

Accessibility is a key concept in land use and transport policies across the world. Several recent reviews have highlighted the huge importance of accessibility for policymakers and society, transport geography and other research areas, and have proposed new research challenges (see Geurs, De Montis, & Reggiani, 2015; Páez, Scott, & Morency, 2012; van Wee, 2016). Although accessibility has been used as an indicator in many spatial scopes and from various perspectives, it is usually used in the context of static studies. In other words, most accessibility studies consider each variable as a single and static value within each study scenario. With regard to accessibility studies, one of the challenges faced by tomorrow's (or even today's) researchers is to improve space-time data input methods (Geurs & van Wee, 2004).

Accessibility is a dynamic attribute of location/person. It truly changes over time due to changes in travel times caused by changes in the transport network (e.g. new transport infrastructure, transport policies and regulations, or the price of transport), to changes in opportunities, e.g. changes in the location and availability of populations and economic activities, or changes in individual mobility capabilities. So, accessibility could be used to measure impacts of dynamic effects on territories, e.g. congestion or changes in levels of service in metropolitan areas. Indeed, congestion assessment, being directly related with evaluating the importance of the robustness, reliability, vulnerability, resilience or flexibility of the transport system (van Wee, 2016), will underpin future accessibility studies. Until recently, dynamic data was very scares but now Information and Communication Technologies (ICT) are greatly facilitating access to it. Companies such as TomTom®, Inrix® or Be-Mobile®, or accessed freely and easily from Here, Google® Maps/Transit, or OpenStreetMaps supply it – some cases by new standardization templates, e.g. Google Transit Feed Specification (GTFS). On the other hand, dynamic data is too big. We can properly use them by Big Data techniques and increased calculation capacity, for example, cloud-computing. These technological improvements have led to a growing field of research involving time-of-day variations in private and public transit accessibility (Geurs et al., 2015)

An increasing number of studies have focused on the effects of time variations in transport system performance on accessibility. However, road congestion studies are still based on static scenario comparison methods. These papers generally consider some peak (unrealistic) and off-peak (legal or free flow conditions) scenarios, and base their comparison of accessibility in these scenarios on the assumption of unvarying road network performance. This methodology has been used to describe the spatial structure of vehicle accessibility to towns and railway stations during peak and off-peak hours in Belgium (Vandenbulcke, Steenberghen, & Thomas, 2009), and the spatial distribution of car travel times in the Greater Toronto Area (Canada) in 4 static design/averaged scenarios (Sweet, Harrison, & Kanaroglou, 2015). Other studies have also been conducted on the effect of including unique congestion values and turn penalties in travel time calculations in Edmonton (Canada) (Yiannakoulias, Bland, & Svenson, 2013).

Studies analyzing public transport, however, do consider variations in the characteristics of the transport network, reflected in the timetables used by different means of transport. Nevertheless, comparisons with car travel times continue to be based on static road networks. For example in the spatial distribution of travel time differences between car and public transport during morning peak and noon conditions in the Flanders region (Belgium) (Dewulf et al., 2015), or in the comparison of accessibility by public transport (with and without changes between lines) and by car in Tel Aviv (Israel) for morning peak and noon scenarios (Benenson, Martens, Rofé, & Kwartler, 2011). Consequently, these simplifications of road networks lead to present the findings using different types of static maps that can only show a snapshot of the calculated situation, thus eliminating interesting information, such as the degree of daily changes in accessibility on the territory.

The aim of this study is to explore dynamic cartographic techniques capable of showing the sequence of variations in car accessibility resulting from speed changes in the road network of London (UK). The map presented in this paper should be as useful in clarifying the impact of congestion in metropolitan areas as those used in other largely time-dependent fields such as meteorology.

The article is divided into the following sections: Section 2 briefly summarizes the different cartographic techniques suitable for the purpose of our study, together with techniques used in earlier studies on accessibility and congestion. Section 3 presents the study variables and the data collection method used. Section 4 details the cartographic techniques used in this analysis. Finally, section 5 outlines our conclusions and proposes lines for future research.

## 2. OLD TECNIQUES IN NEW MAPS

Dynamic accessibility analysis involves analyzing how to use static mapping techniques in dynamic studies without sacrificing soundness and plainness (Bertolini, le Clercq, & Kapoen, 2005). Accessibility is usually mapped using different types of cartographic representations, the most common types being choropleth maps, grids or contour, although other graphically powerful maps have also been used.

Choropleth maps, which use colors to represent accessibility levels by region, are the most widely used in accessibility studies. Obviously, a smaller the size of regions means more accurate results are presented in the map. Several previous studies on accessibility and congestion have presented their results on choropleth maps (see Benenson et al., 2011; Sweet et al., 2015; Vandenbulcke et al., 2009). Grid maps, which are a type of choropleth map, use uniform spaced zones. The advantage of using regular grid data is we could overcome some aspects of the Modifiable Areal Unity Problem (MAUP) (Kwan & Weber, 2008; Spiekermann & Wegener, 1999). Several studies on accessibility have based their data and represented their findings on grid maps(Cheng & Bertolini, 2013; Martin, Wrigley, Barnett, & Roderick, 2002; Páez et al., 2012; Salonen & Toivonen, 2013). The interpretability of both choropleth and grid maps, together with their variations over time, strongly depend on the definition of the value range associated with each color in the legend. These maps can also be extruded to obtain some 3D representations.

Another common method of mapping accessibility or its components involve continuous surfaces created by the interpolation of data points. The most common interpolation maps are Contour maps, also known as heat maps, which are 2D representations of these surfaces. Contour maps use colors and isolines to represent the accessibility value surfaces (Gutiérrez & Urbano, 1996; Ortega, Quintana, & Pastor, 2011; Owen & Levinson, 2014). However, although contour maps are a powerful tool for depicting interpolation results, they are rarely used for 3D maps. In this representation, high accessibility levels appear as 'mountains', whereas 'valleys' represent low accessibility values (see Spiekermann & Wegener, 1996). Gradient colors and isolines also help ensure the intepretability of 2D and 3D surface representations.

Other, less common and used, techniques for representing accessibility findings are cartograms, or anamorphosis maps. This technique is based on distorting geographical areas according to the values to be mapped, thus allowing the results of the accessibility analysis to be presented in a more specific and attractive format. For lineal data, octilinear cartograms simplify the geographical representation of transport networks by representing elements exclusively by horizontal or vertical lines, or 45º angles (Condeço-Melhorado, Christidis, & Dijkstra, 2015). Other interesting techniques are time-space maps, in which elements are organized in such a way that the distances between them are not proportional to their physical distance, but to the travel times between them (Axhausen, Dolci, Fröhlich, Scherer, & Carosio, 2008; Shimizu & Inoue, 2009;

Spiekermann & Wegener, 1994; Ullah & Kraak, 2014). Some of them are presented in animated maps, which have the power to clearly explain the phenomenon studied (ITC - Universiteit Twente, 2011). However, they require higher computational cost algorithms to avoid excessive deformations that would prevent users from recognizing the geographical shapes.

## 3. METHODOLOGY

### 3.1 Dynamic Accessibility Measure

Accessibility is a complex measure with many possible measurement techniques (Geurs & van Wee, 2004). We have used a modified potential accessibility (Hansen, 1959) zone-based indicator to measure the effects of congestion on territorial vehicle accessibility. Its values are obtained from the number and the cost of reaching opportunities from any origin. We can understand the results of this indicator as the sum of accessible opportunities weighted by the value of their impedance according to an impedance-decay function. To study the direct dynamic congestion effect, we just add time variability to the shortest travel time route for each origin-destination relationship and different instances of departure in a network. Opportunity values remain constant. Equation A shows the definition of accessibility and cost estimation

$$A_i^t = \sum_{j \in N} D_j \cdot e^{\beta \cdot c_{ij}^t} \; ; \; \forall \; i \; \in N, t \in T$$

(A)

subject to:

$$c_{ij}^t = \sum_{m \in M} \sum_{e \in E} \alpha_{eij}^{tm} \cdot c_e^m \; ; \; \forall \, ij \in G, t \in T$$

Where:

$A_i^t$ is the potential accessibility value of origin *i,* beginning at instant *t.*

$D_j$ is the opportunities of destination *j*

$e^{-\beta \cdot c_{ij}^t}$ is the impedance-decay function (no-cost trips equal= weight one).

*β* is the parameter. In our case, we used β= -0.065.

$c_{ij}^t$ is the impedance experienced when travelling from origin *i* to destination *j* by the shortest route, beginning at instant *t.* On this paper, the impedance is the travel time [min].

$\alpha_{eij}^{tm}$ is the binary variable that indicates whether network link *e* is used for the trip between origin *i* and destination *j* which has begun at instant *t,* starting at instant *m,*

$c_e^m$ is the expected impedance of network link *e*, use of which begins at instant *m.* In this study, the expected time is the travel time [min]

$N$ represents all the zones included in the calculation area.

$G$ is the set of origin-destination relationships, including relation with itself (origin i = destination j)

$T$ is the set of instants of started trips.

$M$ is all possible instants within the study.

### 3.2    Study Area and Data

The city of London (UK) and its Largest Urban Zone (LUZ) is the most populous and one of the largest metropolitan areas in the EU (ESPON, 2014). In 2013, TomTom® ranked it as the fourteenth most congested urban zone in the whole of Europe (TomTom, 2013).

On this paper, we defined the metropolitan area of London as all LAU1 entities that have more than 50% of their territory within a density isoline of 500 inhabitants/km$^2$ from the main city. We divided these municipalities into a regular grid of 2x2 km (4km$^2$) cells based on the 1 km$^2$ EEA reference grid (Figure 1) (Eurostat, 2006). These were our Origin and Destination zones. We also used all external cells that can be reached via the network from inside the study area within 15 minutes at midnight, to avoid border effects. Downtown cell contains Charing Cross.

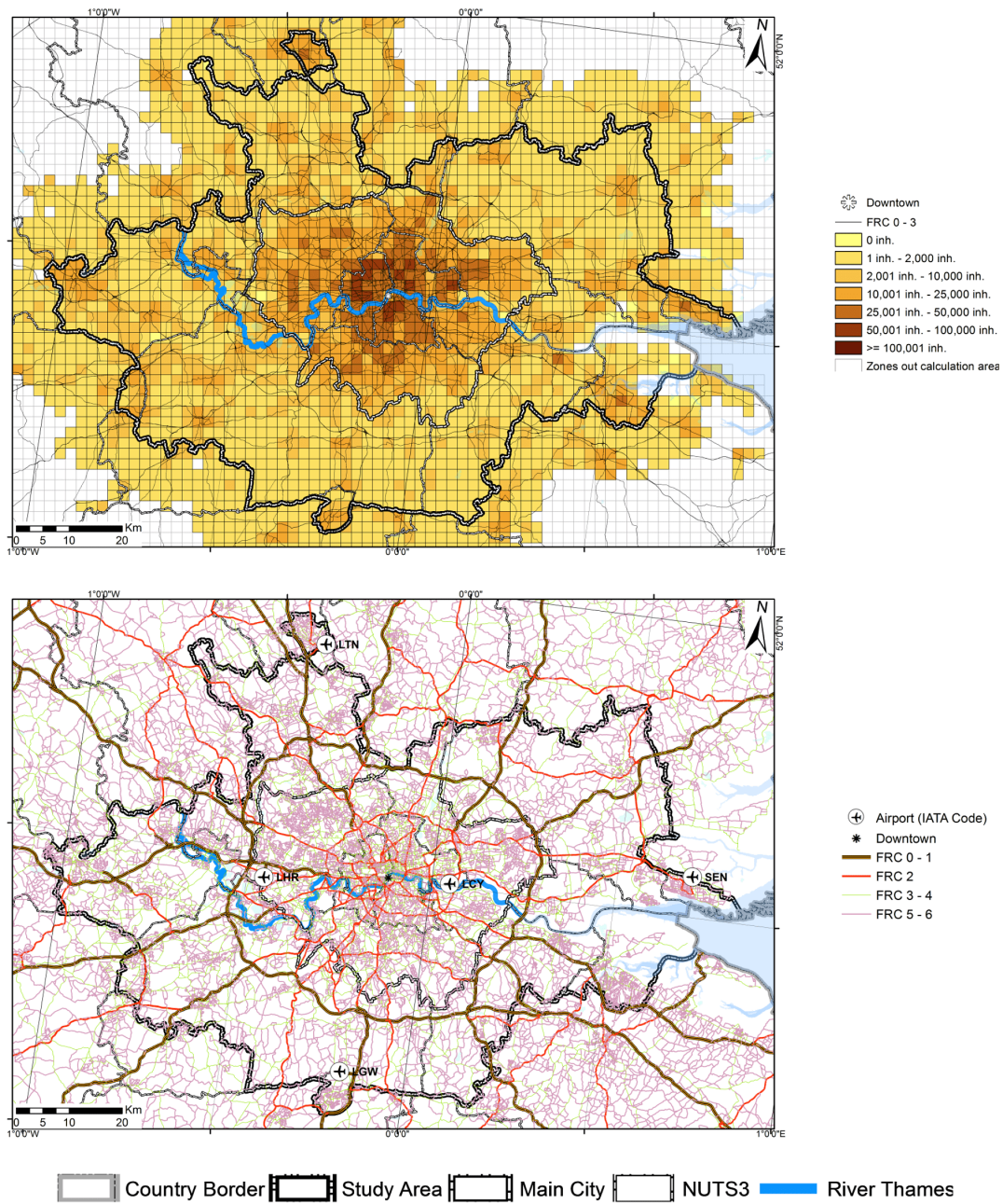

Figure 1. The metropolitan area of London

The transport network was a subset of the TomTom® UK road network (March 2013 version). We only used links in categories 0 to 6 of the Functional Road Classification (FRC). TomTom ® Historical Speed Profile provided network speeds. It is made up of the speed values observed every 5 minutes (TomTom, 2013b), as a percentage of the speed at each moment with reference to the observed free flow speed. In the March 2013 version, speed values between 2011 and 2012 were used. Each link and direction was assigned 1 of the 98 predetermined profiles for each weekday, provided it had a minimum of 1,000 observations every 5 minutes and for each day. The resulting network is fully connected and incorporates all roads and the main urban network. For London Area, our network length is 17,354 km, and 91.10% of drivable lanes, e.g. both direction edges count double, has a Speed Profile, see Figure 2.

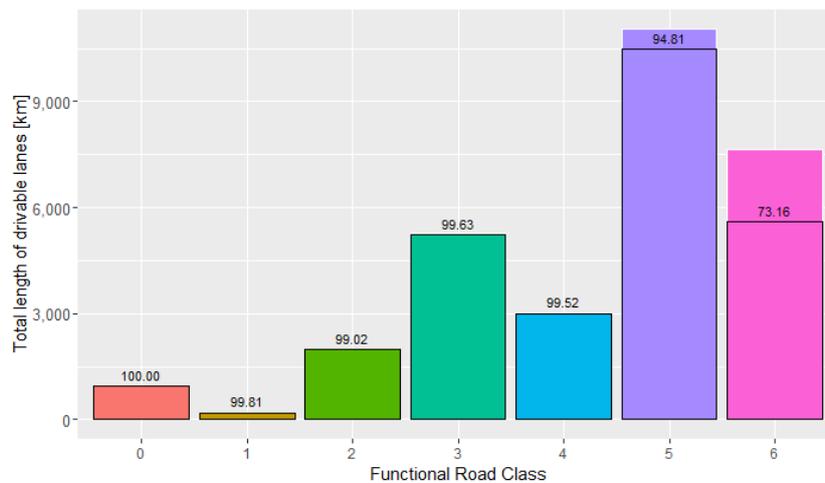

Figure 2. Kilometers of UK road network by category. The back bars are km with speed profiles (showing percentage).

## 4. CONSTRUCTION OF THE MAP

In order to show the dynamic effect of congestion on accessibility in an easily understandable way, we used two types of animations: *a)* isolinear cartograms, and *b)* 3D maps created by extruding the study grid zones. For this purpose, we first generated a series of maps of each scenario, ordered chronologically. These were later combined to form a single animation (each second shows 4 scenarios, i.e., 1 observational hour) The maps have been stripped of all elements that would prevent the observer from focusing on the data of interest.

In this case, the impendence parameter for evaluating accessibility is the travel time between 2 centroids, generating a scenario every 15 minutes. This gives a total of 96 scenarios per day. To facilitate observation, the NUTS3 territorial divisions, the (schematic) study area, the River Thames, the city center and the location of Luton (LTN), London City (LCY), Southend (SEN), Gatwick (LGW), and Heathrow (LHR) airports are the same in all maps and animations (Figure 1).

The first type of animation, created entirely with ArcMap 10.4, is based on isolinear cartograms. These cartograms are a visual representation of the way in which transport distorts geographical perceptions, and how congestion distorts the distortion capacity of transport. As can be seen, a "nearby" location can become "far away" due to congestion (this is the basis of our dynamic accessibility study). For this reason, we have mapped the distortion of territorial divisions, the River Thames and the location of the airports (Figure 3) with respect to a single point (London downtown) for both directions of travel. This representation facilitates interpretation of the effect of congestion on transit by transforming isolines into concentric circles (Figure 3). The map construction process was divided into the following stages (Table 1):

| | | |
|---|---|---|
| Once | Step 1 | Convert each polyline into points, i.e. find their vertices and intermediate points (in our case, maximum allowed distance was 250m). |
| | Step 2 | Obtain the coordinates of all points (previous step result and other point features). |

| | Step 3[a] | Obtain the unit vector between the central point (in our case Downtown) and strep 2 points. |
|---|---|---|
| Once per scenario | Step 4[b] | Interpolate the impedance surface, e.g. using IDW. Interpolating points were Origin/Destination centroids. |
| | Step 5[b] | Assign a module (impedance) according to the previous surface. |
| | Step 6[a] | Obtain the "distorted" position of each point (central point + module · unit vector). |
| | Step 7 | Rebuild/Redraw lines. |

a This step is a sub-step of Ullah and Kraak 's Step I (Ullah & Kraak, 2014)

b We could have avoided this step if we had used distorting points as Origin/Destination points. That strategy would require more computation effort that could have greatly distorted the geography component.

Table 1. Time cartogram construction process

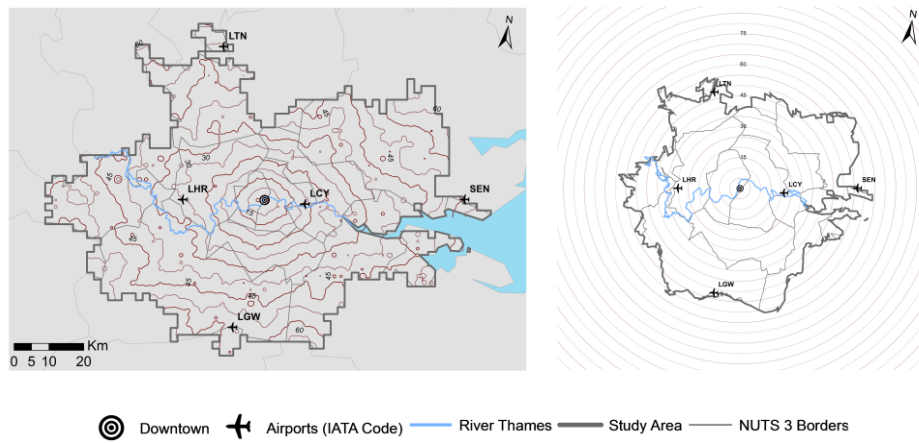

Figure 3. Travel time from Downtown in Free Flow Speed scenario (21:30 – 02:15)

(left: geographic map; right: time cartogram)

Note how this methodology does not prevent point overlapping or overtaking, as their position with respect to the study point only depends on their direction and the impedance value in each scenario.

The London metropolitan area, distorted according to the travel time to and from the city, is seen to "grow" in size as congestion increases travel times throughout the day. Figure 4[1] and Table 2 show changes in the surface area caused by congestion-induced territorial distortion. At peak times, the resulting area is 60% larger than at free flow times.

| Start traveling at | From Downtown | To Downtown |
|---|---|---|
| **08:00** | 127.34 | 162.79 |
| **12:00** | 128.35 | 129.80 |
| **17:00** | 161.08 | 140.28 |
| **22:00** *(21:20 – 02:15)* | 100.00 | 100.00 |

Table 2: Relative time-distorted area

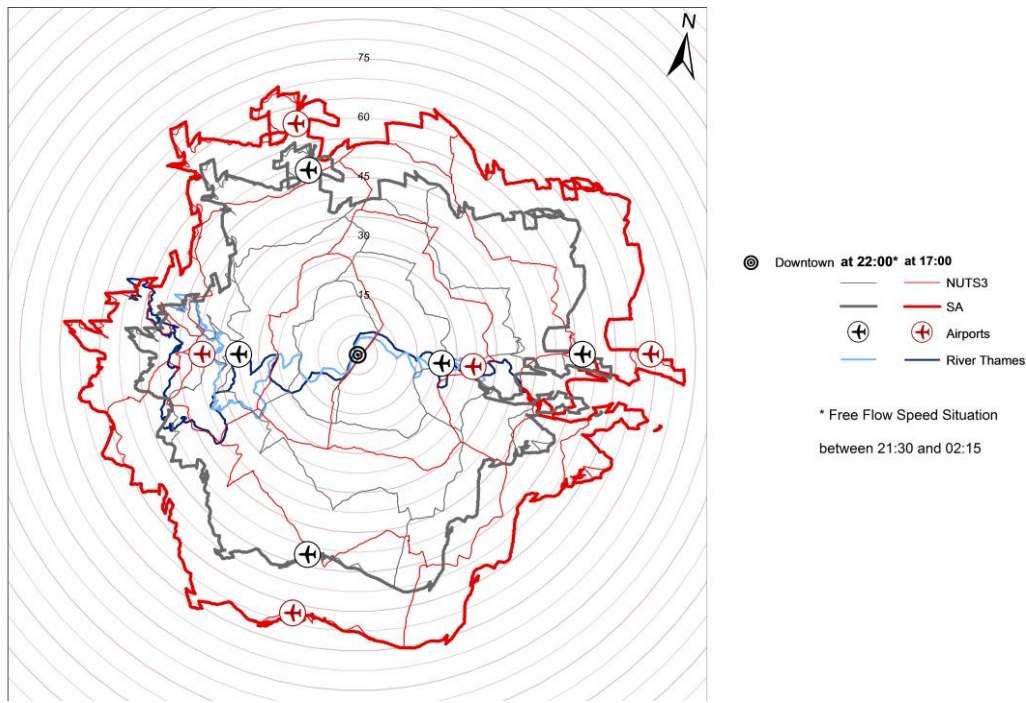

Figure 4. How congestion distorts London and Downtown

The second map sequence is made up by extruding the different accessibility levels obtained in each grid cell and each scenario. The height and color of each column depends on the accessibility level represented (Figure 5). Although accessibility is reduced as congestion increases, dynamic mapping of variations in absolute values can show subtle changes, mask results, or even lead to misinterpretation. For example, we could evaluate a tiny percentage loss as a high loss in any area with initially high values. We avoided this in 2 ways: by showing absolute accessibility values (free flow situation - midnight) in a separate map to be used as a reference, and by mapping congestion-induced accessibility changes (animation) using the same technique, but where height represents the percentage of accessibility in a particular scenario with respect to the same percentages shown in the reference map. In this way, free flow periods are shown with the same height in all zones. The minimum value shown on the main map is 50% of accessibility in free flow periods. This allows us to shows waves of relative loss of accessibility during the day and their distribution over the 24-hour period.

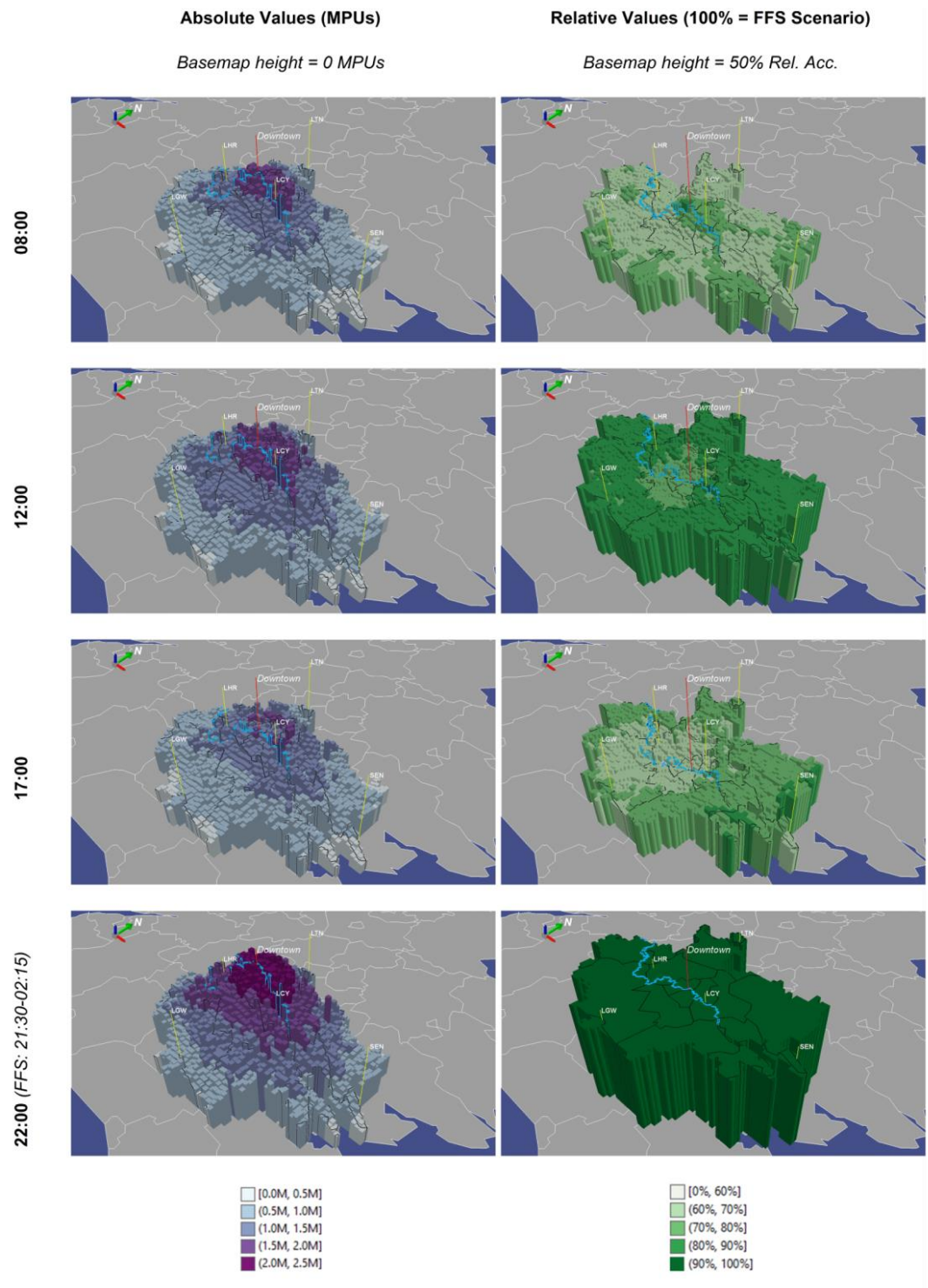

Figure 5. Accessibility in London on Wednesdays

The view shown minimizes the number of zones that could have remained hidden behind other zones due to differences in height. To facilitate understanding, linear elements have also been extruded, and visual continuity is ensured by showing vertical lines on columns whenever required. Point elements have also been mapped.

## 5. CONCLUSIONS AND FUTURE RESEARCH

Access to vast new data sources (Big Data), and the availability of new analysis techniques has improved the scope and accuracy of territorial studies. However, these tools are not without their risks and challenges (Kitchin, 2013), and can call for a review of methodology and hypotheses. Although conventional mapping techniques are still useful and necessary for representing study outcomes, they are sometimes unable to transmit some aspects of the data, and must therefore be adapted to these new techniques. Dynamic accessibility studies, in which findings are not dependent on the status of the road network and/or traffic opportunities at a specific time, but rather on how the network changes thought a trip to reach the destination, are a good example of this need for change.

This study has shown that animation techniques can represent phenomena inherent to dynamic systems, such as trends, which cannot be correctly explained, or can only be partially explained, using static techniques (compare animation and main map). Nevertheless, because of their ephemeral nature, animations should only be used to show information that will clarify the message in the most striking way possible, for example, using colors, height and distortion. Any stationary element that could undermine the simplicity of the animation should only be represented on a static map. The combination of animations and static maps used in this study clearly explains the uneven distribution of the impact of congestion in terms of both time and geography. This technique shows which areas are affected by congestion, rather than the network arcs where congestion occurs.

Finally, our maps still follow the paradigm of showing what the author wants to show. We have not explored other, and potentially highly useful, representations of our results that involve a certain degree of user interaction, such as the zone on which the dynamic cartograms are focused, or the 3D view of the data. This interactive presentation could potentially be resolved following web-based models.

## 6. SOFTWARE

The maps using in this study were created with ArcGIS 10.4: ArcMap was used to create the cartograms, and ArcScene to create 3D maps. Travel times between origin and destination zones were generated by the OD Cost Matrix feature of the Network Analyst tool. Various customized Python scripts were generated to calculate accessibility based on the cost matrix, to automatically export and adjust the maps included in the animations. The animations were created using Photoscape and the main map was created using Adobe Illustrator CS6.

## ACKNOWLEDGEMENTS

The authors would like to thank the Spanish Ministry of Economy and Competitiveness for funding this research as part of the SPILLTRANS (TRA2011-27095) and DynAccess (TRA2015-65283-R) projects.

**MAP DESIGN**

The main map is divided into 2 parts. The first shows the data used to calculate dynamic accessibility levels. The maps were created using the LAMBERT AZIMUTAL EQUAL AREA (LAEA) projection – Datum ETRS_1989 for Europe, using an extension that showed the entire area of study at a scale of 1:500,000 on paper size DIN A-4. These were used to show the elements needed to help the observer interpret the cartograms and 3D maps (territorial divisions, the River Thames, the airports and downtown) to prevent information overlap, and to create a reference grid with main 1-degree lines and secondary half-degree lines. The second part, which is the landscape view of the map, includes a series of cartogram (created from unit vectors derived from the foregoing maps) and extruded 3D map sequences, also created using the above projection. 19 snapshots were chosen to represent major changes in accessibility, based on the overall trends in accessibility show in the top chart of the Main Map.